# Element- and momentum-resolved electronic structure of the dilute magnetic semiconductor manganese doped gallium arsenide

Slavomír Nemšák[1,2,3,*,†], Mathias Gehlmann[1,2,3], Cheng-Tai Kuo[1,2], Shih-Chieh Lin[1,2], Christoph Schlueter[4,‡], Ewa Mlynczak[3], Tien-Lin Lee[4], Lukasz Plucinski[3], Hubert Ebert[5], Igor Di Marco[6,7], Ján Minár[8], Claus M. Schneider[1,3], Charles S. Fadley[1,2]

[1]*Department of Physics, University of California, 1 Shields Ave, Davis, CA-95616, USA*

[2]*Materials Sciences Division, Lawrence Berkeley National Laboratory, 1 Cyclotron Rd, Berkeley, CA-94720, USA*

[3]*Peter-Grünberg-Institut PGI-6, Forschungszentrum Jülich, 52425 Jülich, Germany*

[4]*Diamond Light Source, Harwell Science and Innovation Campus, Didcot OX11 0DE, UK*

[5]*Department of Chemistry, Ludwig Maximillian University, D-81377 Munich, Germany*

[6]*Department of Physics and Astronomy, Uppsala University, Box 516, SE-75120 Uppsala, Sweden*

[7]*Asia Pacific Center for Theoretical Physics, Pohang 37673, Republic of Korea*

[8]*New Technologies-Research Center, University of West Bohemia, 306 14 Plzen, Czech Republic*

**Abstract**

The dilute magnetic semiconductors have promise in spin-based electronics applications due to their potential for ferromagnetic order at room temperature, and various unique switching and spin-dependent conductivity properties. However, the precise mechanism by which the transition-metal doping produces ferromagnetism has been controversial. Here we have studied a dilute magnetic semiconductor (5% manganese-doped gallium arsenide) with Bragg-reflection standing-wave hard X-ray angle-resolved photoemission spectroscopy, and resolved its electronic structure into element- and momentum- resolved components. The measured valence band intensities have been projected into element-resolved components using analogous energy scans of Ga *3d*, Mn *2p*, and As *3d* core levels, with results in excellent agreement

[*] Correspondence to: snemsak@lbl.gov
[†] Current address: Advanced Light Source, Lawrence Berkeley National Laboratory, 1 Cyclotron Rd, Berkeley, CA-94720, USA
[‡] Current address: DESY Photon Science, Deutsches Elektronen-Synchrotron, 22603 Hamburg, Germany

with element-projected Bloch spectral functions and clarification of the electronic structure of this prototypical material. This technique should be broadly applicable to other multi-element materials.

**Introduction**

The dilute magnetic semiconductors show great promise in spintronic applications, due to their ferromagnetic order at relatively high temperatures, and the potential for such devices as spin-current based transistors or lasers [1]. However, the origin of the ferromagnetism in the III-V dilute magnetic semiconductors has generated controversy since one of the first prototypical materials, manganese-doped GaAs, was discovered [2], with detailed discussions appearing elsewhere [3,4,5,6]. For example, the double exchange mechanism, in which the Mn *3d* electrons form an impurity band separate from the GaAs *4p*-bands, or the *p-d* exchange mechanism, where Mn-As hybridization occurs, are two scenarios widely discussed. It is now generally agreed that both mechanisms have to be taken into account in a unified model, a conclusion supported by prior results from more bulk sensitive hard X-ray angle-resolved photoemission (HARPES) on $Ga_{1-x}Mn_xAs$ and GaAs [7,8], resonant soft-ray angle-resolved photoemission spectroscopy (ARPES) on $Ga_{1-x}Mn_xAs$ [9,10], and most recently very low energy ARPES on $Ga_{1-x}Mn_xAs$ [11].

Disentangling the character of the electronic valence states of a material requires two things: bulk sensitivity, which suppresses the surface and contaminant contributions, and a method for site-specific identification. The first can be achieved by increasing the energy of the exciting radiation to the hard/tender X-ray regime of a few keV, which enhances the information depth (controlled by the inelastic mean free path of the photoelectrons) to tens of nm [7,8]. The second can be achieved by resonant excitation of ARPES [9,10], but this involves many-electron effects including the core hole that prevent a quantitative analysis, or by making use of Cooper minima in ARPES [12], but these are more subtle effects, which are limited to only certain elements and subshells. We believe this second goal is more universally and quantitatively achieved by the combination of HARPES and standing wave (SW) excitation, as the spatial phase information provided by the SW, coupled with core level intensities, permits a much more quantitative decomposition of the VB intensities into their atomic components.

In particular, standing-wave excitation from Bragg crystal planes provides a way to differentiate a signal from the specific sites within a single unit cell of a single-crystalline or epitaxial material. This has previously been successfully employed by Woicik et al. [13,14], Kim et al. [15] and Thiess et al. [16,17] to decompose valence-band spectra into site-specific components. However, the momentum information was not accessible in these experiments due to some combination of the lack of angle-resolving capabilities in the energy analyzers used and the momentum-smearing effects of phonon excitations, which significantly lower the fraction of direct transitions in hard X-ray photoemission [7,8,18]. Nevertheless, through cryogenic cooling to suppress phonon effects, and with sufficient angular resolution in detection, such experiments should also be possible in an angle-resolved mode, as we shall demonstrate.

Here we present a HARPES study of Mn-doped and pure undoped GaAs that goes significantly beyond prior works in involving SW excitation. Such SW-HARPES data permit determining the electronic structure with unprecedented element/site- and momentum- resolution. The experimental results are compared with density functional theory, as implemented in the Korringa–Kohn–Rostoker (KKR) method [19,20].These calculations are coupled via **k**-conserving transitions to a free-electron final-state (FEFS), and the data are interpreted through Bloch-spectral functions (BSF) projected onto different sites in the unit cell.

**Results**

We now discuss some proof-of-principle SW-HARPES results for the same two systems of GaAs(001) and a thin film of the DMS $Ga_{1-x}Mn_xAs$ (001) grown on GaAs(001), using a hard X-ray photoemission facility particularly well suited for this at the Diamond Light Source Beamline I09 [21]. In these measurements, the incident X-ray geometry was set to excite a (-1-1-1) reflection, and the photon energy of around 2.7 keV was scanned over the Bragg condition. During this scan, the resulting SW moves by one half of its period, which is the spacing of the (-1-1-1) planes, along the direction of the reciprocal lattice vector **g**$_{-1-1-1}$ [13,14]. In Fig. 1(a), we illustrate the diffraction geometry used, with the SW scanned along the [111] direction, and four possible positions of its antinodes indicated. Fig. 1(b) schematically indicates the relationship of the SW to the (111) planes in the crystal. More details on the experimental geometry and measurement procedures are in the Methods section. Fig. 2 shows the experimental geometry, with various key parameters indicated.

**Core-level standing-wave photoemission data**

In Fig. 3, photon-energy scans of different photoemission core-level peak intensities are shown. In Fig. 3(a) we report scans for undoped GaAs, which are in good agreement with prior core-level data of this type for GaAs(111) by Woicik et al. [14]. In Fig. 3(b) we report scans for a sample consisting of a film of $Ga_{0.95}Mn_{0.05}As$ deposited on a GaAs(001) substrate with a thickness $D_{film} \approx 100$ nm. Figs. 3(a) and 3(b) show that energy scans of Ga *3d* and As *3d* are clearly very different for both samples, due to their different positions in the unit cell along the [111] direction. In Fig. 3(b) one can also see two separate Bragg peaks: a narrower one corresponding to the semi-infinite GaAs substrate and a wider one to the finite-thickness $Ga_{0.95}Mn_{0.05}As$ (note the difference in scale of these two panels). It is also important to stress here that the inelastic mean free path of electrons with kinetic energy of ~2.7 keV is only around 5 nm, so all the detected core-level photoelectrons originate from the top ~20 nm of much thicker $Ga_{0.95}Mn_{0.05}As$ film. The shift in energy of the wider $Ga_{0.95}Mn_{0.05}As$ Bragg peak is due to the fact that the Mn doped film has a ~0.25% larger lattice constant (5.674 vs 5.660 Å [1,20]) and hence the Bragg condition at the same incidence angle is fulfilled for a larger photon wavelength (lower photon energy side). This additional peak is also much broader, since the number of effective scattering layers is lowered compared to the substrate. More quantitatively, based on the dynamical diffraction theory of x-rays the extinction depth of the GaAs(111) reflection at 2.718 keV is about 300 nm, which represents the minimum thickness for a (111)-oriented semi-infinite GaAs slab to fully diffract x-rays with the (111) reflection at this energy. For a (001)-oriented GaAs film the corresponding (111) extinction depth becomes 173 nm, simply due to the angle of exit from the surface, and this is still significantly larger than the 100 nm $Ga_{0.95}Mn_{0.05}As$ film thickness. Thus, full multilayer diffraction is not possible in the film, and this results in the broadening of the secondary Bragg peak in the energy scans (Figure 3(b)). So, for the $Ga_{0.95}Mn_{0.05}As$ sample, there are really two overlapping and interfering SWs moving along the [111] direction as energy is scanned, but the actual photoelectrons arise only from the DMS film. Fig. 3(c) confirms this interpretation, showing the results of dynamical diffraction calculations for this sample, which agree excellently with experiment.

It also can be easily recognizable how the sample orientation affects the position of these respective Bragg peaks. Apparent Bragg energies for scans shown in Figures 3(a), (b) and (d) are all slightly shifted with respect to each other due to the slight misalignment in azimuthal ($\Phi$) and tilt angles. As we discuss later, this misalignment, which cannot be corrected for without having a full 6-axis goniometer, is also responsible for mapping slightly different regions in the Brillouin zones of the two samples. This technical aspect, however, does not affect the conclusions provided in this work, as all the theoretical calculations were performed using custom trajectories in reciprocal space derived by comparing experiment to free-electron final-state calculations, as done previously [7,8].

Figure 3(d) shows similar experimental energy scans for the doped sample for three core levels: Ga 3$d$, As 3$d$, and Mn 2$p$, but with a larger energy step. Although the scan for Mn 2$p$ is noisier due to the low concentration of Mn, the Mn curve agrees within statistical error with that for Ga, suggesting a predominantly substitutional position of the Mn atoms. The same conclusion has been derived from a similar DMS sample using a SW created with (311) reflection, where the Ga and Mn scans are essentially identical (data not shown here). These two sets of data confirm a high degree of on-site substitution, rather that interstitial doping, in our samples. This is in good agreement with previous X-ray SW work, which revealed the dependence of Mn dopant site occupancy in GaAs as a function of its concentration [22]. It is, however, encouraging that this ordering and substitutional location of Mn is preserved even at the relatively thin surface region as studied here by photoemission.

**Angle-resolved valence-band photoemission data**

We now add in a stepwise fashion elemental sensitivity to HARPES via such SW measurements in Fig. 4. In Figures (a) and (b) we show HARPES results for GaAs and $Ga_{0.95}Mn_{0.05}As$, over the acceptance angle of 25 degrees of the Scienta analyser used. These are shown for a representative photon energy of 2719 eV. It is important to note here that changes in $k_z$ and overall probed trajectories in the reciprocal space are relatively small with a small change of excitation energy that has been used in this experiment. Specifically, $k_z$ for valence photoelectrons collected along the sample normal only changes by ~0.05 Å$^{-1}$ as we scan photon energy from 2713 to 2725 eV. These changes are negligible comparing to the size of the Brillouin zone, which is 2.2 Å$^{-1}$ along Γ–X–Γ direction. These conclusions are graphically illustrated in Supplementary Figure 1.

Overlaid on these data are light gray curves calculated for direct transitions from the ground-state electronic structures of both materials, with the coherent potential model being used for the doped material, to a strictly free-electron final state (FEFS). Such FEFS calculations provide a very useful method for initial analysis of HARPES data [7,8] permitting the determination of the exact orientation of the sample (Gehlmann M. & Plucinski, L. code for calculation of free-electron final-state ARPES, to be published). This method is accurate to less than 0.5°, including a small tilt of each sample for the present data and revealing a slight difference in geometry between the two samples. These calculations also result in a set of initial **k** points, or really a path in **k**, that is measured by the detector, and along which subsequent element-resolved Bloch spectral functions will be determined. The precise paths used in our analysis are shown in Supplementary Fig. 1.

**Discussion**

In Figures 4(c) and (d), we show the experimental data again, but after a procedure making use of the core-level energy scans for Ga 3*d* and As 3*d* that permits identifying the element-resolved contributions to each HARPES pixel. This method proceeds in the following way. We first assume that the intensity in each pixel at binding energy $E_B$, wave vector **k**, and photon energy $h\nu$, $I_{HARPES}(E_B,\mathbf{k},h\nu)$, can be described as a superposition of a contribution from As, $I_{As}(E_B,\mathbf{k},h\nu)$, and from Ga (or Mn), $I_{Ga(Mn)}(E_B,\mathbf{k},h\nu)$, as:

$$I_{VB}(E_B, \mathbf{k}, h\nu) = I_{As}(E_B, \mathbf{k}, h\nu) + I_{Ga}(E_B, \mathbf{k}, h\nu) + I_{Mn}(E_B, \mathbf{k}, h\nu). \quad [1]$$

It has long been realized through various theoretical studies and experimental comparisons that, as the photon energy is increased, the matrix elements and cross sections for valence photoemission are increasingly controlled by the inner spatial regions of each atom involved [23,24,25], thus making the decomposition above a better and better approximation. This sort of superposition has also been used previously to analyze SW spectra in the hard X-ray DOS limit [13,14,15,16,17], but not with momentum resolution, which is unique to this study.

In the next step, we assume that the SW variation of these components with photon energy will follow the core intensity for a given atom as a function of photon energy, again an assumption that has been used before [13-17]. Each core intensity is normalized to unity away from the Bragg reflection, such that it should represent the fractional variation of a given component of the valence intensity as the SW position moves. Because Ga and Mn occupy the same site type, we cannot distinguish them in this superposition. Then, the energy-dependence of each pixel is projected into fractional As and Ga+Mn components $f_{As}$ and $f_{Ga+Mn} = 1 - f_{As}$ by using a least-squares comparison to the core-level intensities:

$$I_{VB}(E_B, \mathbf{k}, h\nu) = f_{Ga+Mn}(E_B, \mathbf{k})\bar{I}_{Ga3d}(h\nu) + f_{As}(E_B, \mathbf{k})\bar{I}_{As3d}(h\nu), \quad [2]$$

with $\bar{I}_{Ga3d}(h\nu)$ and $\bar{I}_{As3d}(h\nu)$ being the energy-scans shown in Figs. 3(a) and 3(b), normalized to 1 off the Bragg reflection, and adjusted in amplitude according the differential cross section ratios

$\frac{d\sigma_{Ga4s,4p}}{d\Omega} / \frac{d\sigma_{Ga3d}}{d\Omega}$ and $\frac{d\sigma_{As4s,4p}}{d\Omega} / \frac{d\sigma_{As3d}}{d\Omega}$. $\bar{I}_{Ga3d}(h\nu)$ and $\bar{I}_{As3d}(h\nu)$ are thus used as basis functions for this projection. The resulting fractions in Eq. 2 are then multiplied by the raw data $I_{VB}(E_B,\mathbf{k},h\nu)$ at each pixel to arrive at an element-projected image, with the contributions from Ga and Mn being implicitly summed.

Once the two $f$ quantities have been determined, a color scale going from +1.0 = maximum Ga+Mn/minimum As to -1.0 = maximum As/minimum Ga+Mn is applied to the raw data at every pixel. The results are shown in Figures. 4(c) and 4(d), and they reveal a dramatic difference between the top and bottom sets of bands, which show strong As character and the middle bands, which are strongly Ga+Mn-like.  For the doped sample, one can even see evidence of what is probably weak Mn intensity in the upper bands over 0-6 eV, but particularly near the Fermi level, that is consistent with prior conclusions that Mn affects the entire band structure [8,9,10]. As an alternate method of looking at the elemental decomposition, Supplementary Figure 2 shows the two different components As and Ga / Ga+Mn separately.  In summary, using the above-mentioned decomposition technique based on SW excitation in the valence band data and the projection onto Ga *3d* and As *3d* core-level intensity modulations, we have obtained an element- and **k** resolved band structure, derived directly from experiment.

The purely experimental quantities derivable from Eq. 1 can also, through the same assumptions of prior work [13-17, 23-25] be related to partial densities of states $\rho_{As}$ and $\rho_{Ga}$ and the differential photoelectric cross sections of the dominant orbitals making up the valence states, here being Ga *$4s^2 +4p^1$*, As *$4s^2+4p^3$*, and Mn *$3d^5+4s^2$*, as averaged over the spectrometer detection range:

$$I_{As}(E_B, \mathbf{k}, h\nu) \approx \rho_{As}\langle d\sigma_{As4s+4p}/d\Omega \rangle \quad [3]$$

$$I_{Ga+Mn}(E_B, \mathbf{k}, h\nu) \approx \rho_{Ga}\langle d\sigma_{Ga4s+4p}/d\Omega \rangle + \rho_{Mn}\langle d\sigma_{Mn3d+4s}/d\Omega \rangle \quad [4]$$

Implicit here is that the photoelectric cross-sections of the orbitals involved are independent of the binding energy and the photon energy within their respective ranges. This simplification is justified by the fact that the photon energy is varied only over ca. 0.25% and the angular range of the acquired data is only 25°. The differential cross-section values at $hv = 3.0$ keV are then used, with their respective values of $d\sigma_{As\ 4s+4p}/d\Omega = 0.4225$ and $d\sigma_{Ga\ 4s+4p}/d\Omega = 0.2345$ [26] and Mn being neglected in what follows due to the small dopant concentration.

Theoretical calculations also confirm the accuracy of our decomposition method, as shown in Figures 4(e) and 4(f). Here, we show element-resolved Bloch spectral functions (BSF), computed for the same trajectories in **k**-space derived by fitting the data in Figs. 4(a) and 4(b) with FEFS calculations. The coherent potential approximation (CPA) method has been used for the doped material, and is shown in the same color scale. There is excellent agreement as to the major elemental ingredients in each band, including significant Ga+Mn near the Fermi level. We do not expect fully quantitative agreement, as more sophisticate inclusion of matrix elements or cross sections would be needed for a direct comparison between experiment and theoretical calculations. Such data should in the future provide a detailed test of both ground-state electronic structure theory and one-step photoemission calculations.

It is also significant that, in order to achieve this agreement between theoretical BSFs and experimental results, we had to treat correlation effects for the Mn $3d$ states with high accuracy. This was achieved by means of the LSDA+DMFT method (cf. Methods section), which predicts Mn $3d$ states extending over a broad range of energies [10]. More simplified treatments of the electron-electron interaction are not able to describe these spectral features. For example, the LSDA+U approach predicts that most Mn $3d$ states form a main peak of limited width (2 eV), with its exact location depending on the value of U. This scenario clearly fails to describe the experimental data reported in Fig. 4(d). In addition, several weak flat-band localized or density-of-states like features at binding energies of ca. 3-5 eV are present in the LSDA+DMFT calculations of Fig. 4(f), but are not seen in experiment. These features have until now been seen only in resonant photoemission [10,27], and may be due to the disruption of long-range order in the doped material.

As a final alternative method for comparing experiment and theory, the equivalent element-specific decomposition analysis was performed on our data at what should lead to the MEWDOS limit, obtained approximately from experiment by integrating over the 25° detector angle. These respective Ga+Mn and As components were then compared with normal theoretical DOS projections. Figures 5(a) and 5(b) contain such data for both undoped and doped samples, together with their theoretical counterparts. The spectra are aligned and normalized to the Ga strength of the peak at ~7.8 eV for the undoped sample and ~7.0 eV for the doped sample. Panel (b) also contains a DOS projection onto Mn, revealing the $E_B$ intervals where one could expect maximum contributions of Mn to the spectral weight. The agreement between the experimental and theoretical projections is again remarkable, including the details of the spectral shape of all three major peaks in the valence band spectra and their relative intensities. Only the highest-binding-energy peak at ~12-12.5 eV is broader in experiment, and shifted slightly to higher energy; these states are strongly As in character and rather localized, and can decay by Auger processes to yield shorter lifetimes. We attribute the greater widths to these lifetime effects, and perhaps also to greater phonon smearing due to the more localized nature of these states.

In conclusion, we have used standing wave excitation produced by the (-1-1-1) Bragg reflection from a GaAs(001) single-crystal and an epitaxial thin film of $Ga_{0.95}Mn_{0.05}As$ to obtain element-, spatial- and momentum- resolved information on the electronic structure of this prototypical dilute magnetic semiconductor. The photon-energy dependence of the Ga *3d*, As *3d* and Mn *2p* core levels over the Bragg condition shows that the Mn dopants are substitutional and occupy Ga sites. The energy difference of the Bragg peaks for the undoped GaAs substrate and the Mn-doped thin film corresponds to the difference in their respective lattice constants. Finally and most significantly, a method for decomposing the HARPES results into their element- and momentum- resolved components is introduced, and the results obtained are in an excellent agreement with site-projected Bloch spectral functions, overall confirming the influence of Mn doping throughout the band structure. Angle-averaging the HARPES results is also found to give excellent agreement with element-projected densities of states.

We thus view these results as most positive for the future use of SW-HARPES for studying the element- and momentum- resolved bulk electronic structure of multicomponent materials. Making use of simultaneous analysis of multiple Bragg reflections, as in prior MEWDOS-level studies [13,14,16,17], should improve the accuracy of localizing the electronic structure, both in momentum and within the unit cell.

**Methods**

**Sample synthesis and measurements**

The sample of $Ga_{0.95}Mn_{0.05}As$ were grown according to an established method in the group of H. Ohno [2,28]. A 100 nm thin film of $Ga_{0.95}Mn_{0.05}As$ was prepared by molecular beam epitaxy onto a GaAs(001) single-crystalline substrate. GaAs was studied as a single crystal. In order to get rid of the surface oxidation and other contaminants, both doped and undoped samples were etched in concentrated HCl for 10 minutes just before their introduction to the UHV chamber. In order to minimize the momentum-smearing effects of phonons on the data, the HARPES measurements were performed with samples cooled to 20 K, which is also below Curie temperature of the DMS (~60 K). The experimental geometry was optimized for (-1-1-1) reflection excitation; in particular, the incident photon beam was incident on the sample at 14° with respect to surface, along the [100] surface direction (see Fig. 2). This was achieved by using 5-axis goniometer at the I09 Beamline of the Diamond Light Source [21], where also our photoemission measurements took place using a wide-angle high-energy hemispherical electron analyzer VG Scienta EW4000. The combined energy resolution (beamline and analyzer) was better than 300 meV, the angular resolution was 0.3°. The energy axes in Figures 4 and 5 were calibrated using valence band maxima of both samples. For undoped GaAs, Fermi level was placed in the middle of the calculated band gap; Ga(Mn)As sample was considered as metallic.

**Theoretical calculations**

Electronic structure calculations were performed with the spin-polarized relativistic Korringa-Kohn-Rostoker (KKR) method [19]. Ground state calculations of substitutionally disordered $Ga_{0.95}Mn_{0.05}As$ have been performed by means of the coherent potential approximation alloy theory. To include electronic correlations, the charge and self-energy self-consistent LSDA+DMFT scheme was employed, which is also based on the KKR approach [20]. LSDA+DMFT calculations were performed for the paramagnetic phase at $T$=400K. Although the sample temperature during our measurements was below $T_C$, and it might be expected that we would carry out calculations for the ferromagnetic state, from a prior resonant photoemission study [27], the actual changes in band positions from the paramagnetic to ferromagnetic are only of the order of 40 meV, and so are not expected to be resolvable with our ~300 meV energy resolution. In addition, carrying out LSDA+DMFT calculations for the ferromagnetic state would require dealing with the so-called double-counting term, which may introduce a substantial error in spin-polarized LSDA+DMFT calculations [29]. For the paramagnetic state, then, we used a Hubbard parameter $U$= 6 eV and the double counting was considered in the fully localized limit. The impurity problem arising in DMFT was solved by the exact diagonalisation technique, using 10 Mn *3d* orbitals and 22 effective bath sites. For more technical details on the solution of the effective impurity model, see Ref. 27. The sample alignment and the path in momentum space sensed by our detector was determined by calculating ground-state band structures for both GaAs and $Ga_{0.95}Mn_{0.05}As$ using the Wien2k code and coupling them via direct transitions to free-electron final states, a method that has been used successfully in prior soft- and hard- X-ray ARPES studies (refs. 7,8 and earlier refs. therein), using a specially written code. By fitting these calculations to experiment (cf. Figs. 4(a) and 4(b)) an optimized **k**-path has been used to calculate LSDA+DMFT band structures, which are finally represented in the terms of element-projected Bloch spectral functions (Figures 4(e) and 4(f)), or as projected densities of states (Figure 5).

**Data availability**

The datasets generated during and/or analysed during the current study are available from the corresponding author on reasonable request.

**Acknowledgements**

The assistance of Sven Döring and Markus Eschbach in carrying out these experiments is gratefully acknowledged.  C.S.F. has been supported for salary by the Director, Office of Science, Office of Basic Energy Sciences (BSE), Materials Sciences and Engineering (MSE) Division, of the U.S. Department of Energy under Contract No. DE-AC02-05CH11231, through the Laboratory Directed Research and Development Program of Lawrence Berkeley National Laboratory, through a DOE BES MSE grant at the University of California Davis from the X-Ray Scattering Program under Contract DE-SC0014697, and through the APTCOM Project, "Laboratoire d'Excellence Physics Atom Light Matter" (LabEx PALM) overseen by the French National Research Agency (ANR) as part of the "Investissements d'Avenir" program. J.M. would like to thank CEDAMNF project financed by Ministry of Education, Youth and Sports of Czech Rep., project No. CZ.02.1.01/0.0/0.0/15.003/0000358.

**Authors' contributions**

SN, MG, CTK, SCL, CS, EM, TLL, CSF performed the experiments. SN, MG, CTK analyzed the data. MG, LP, HE, IDM, JM provided theoretical calculations. CMS, CSF supervised the research. SN, CSF wrote the paper. All authors contributed in the discussions.

The authors declare no Competing Financial or Non-Financial Interests.

**Figure 1**: Schematics of the experiment. (a) The unit cell of undoped GaAs with four different (111) planar cuts intersecting different sites of its zinc blende structure. A {111}-type reflection, actually (-1-1-1), was chosen because the Ga(Mn) and As atoms do not lie together on these planes, with the type of atom intersected by each plane indicated in the figure. (b) A schematic side view of the experimental geometry, where the oscillations of the excitation field along the [111] direction are depicted and different atomic species are excited with a different intensity of primary radiation depending on the position of the nodes and antinodes of the standing wave.

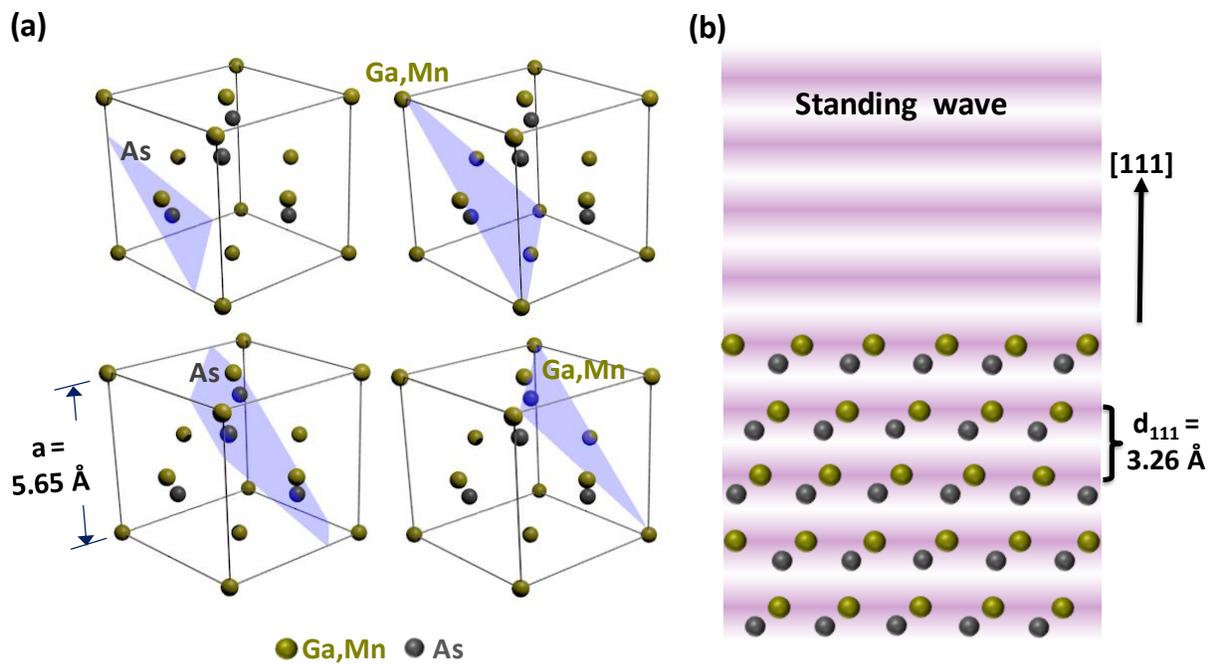

**Figure 2:** The experimental geometry with various key parameters shown. Azimuthal angle $\Phi$, polar and beam incidence angles $\Theta$, $\Theta_{inc}$ and tilt angle are indicated. Light scattering plane runs along [100] crystallographic direction.

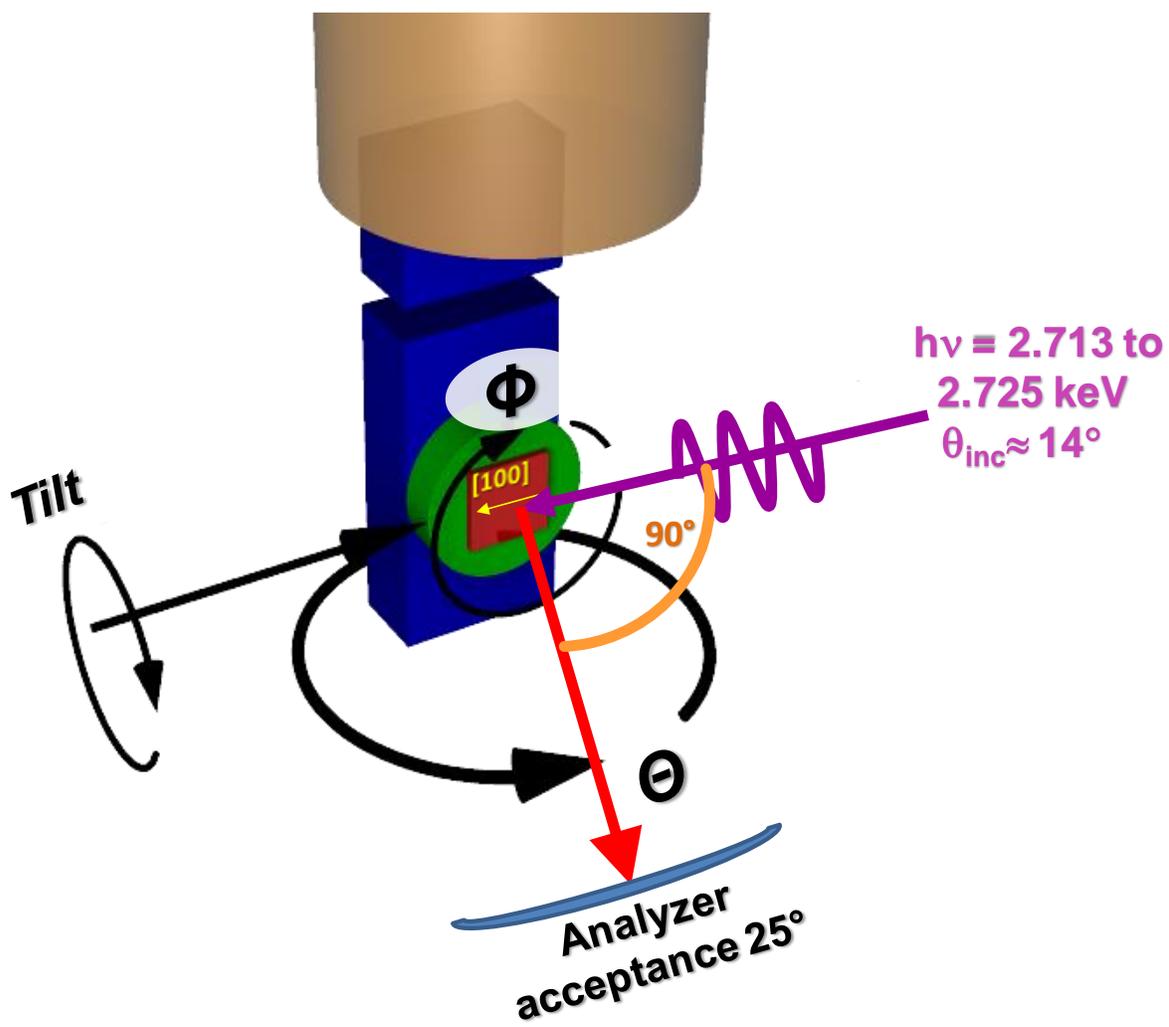

**Figure 3:** Energy-scan standing-wave core-level intensities. The photon energy was scanned over the (-1-1-1) Bragg condition, centered at about 2,718 eV. (a) The variation of As *3d* and Ga *3d* intensities from undoped GaAs through such an energy scan, with obvious strong differences in behavior. (b) A similar energy scan for a sample with 100 nm of $Ga_{0.95}Mn_{0.05}As$ grown on GaAs(001). Two Bragg peaks are observed, due to the slight difference in the lattice constant of the doped sample, with similar differences in phase between As *3d* and Ga *3d* in going over both of the peaks. (c) Dynamical diffraction simulations of the photoelectron intensity from different sites as a function of photon energy, which show very good agreement with experiment in panel (b). (d) An energy scan of Ga *3d*, As *3d* and Mn *2p* over narrower range for the doped sample. The near identity of the curves for Ga and Mn indicates a high degree of substitutional sites for Mn, and was confirmed with similar data for a (311) reflection (not shown).

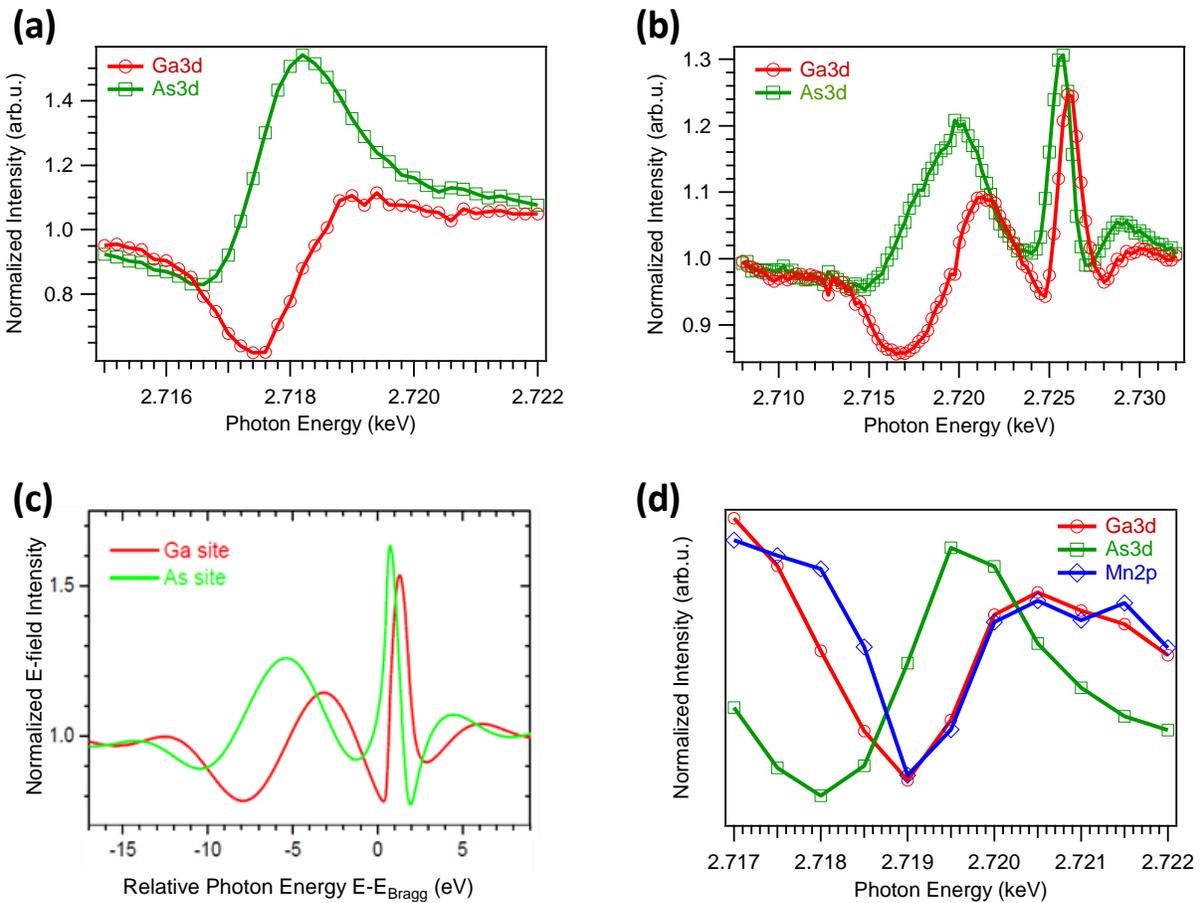

**Figure 4:** Experimental hard X-ray angle-resolved photoemission data. Data shown for (a) GaAs and (b) (Ga,Mn)As in a (-1-1-1) reflection geometry, with photon energy of ~2.7 keV. The light gray curves are calculations based on a local-density band structure excited by direct transitions into free-electron final-states, and were used to determine the precise sampling in **k**–space, which was slightly different for the two samples. (c),(d) Experimental decomposition into Ga+Mn and As components using core-level intensities and Equations [1] and [2]. The color scale is maximum Ga,Mn = 1.0 and maximum As = -1.0. (e),(f) theoretical calculations of element-resolved Bloch spectral functions, using the coherent potential approximation for (Ga,Mn)As and the same sampling in **k**–space as (a) and (b).

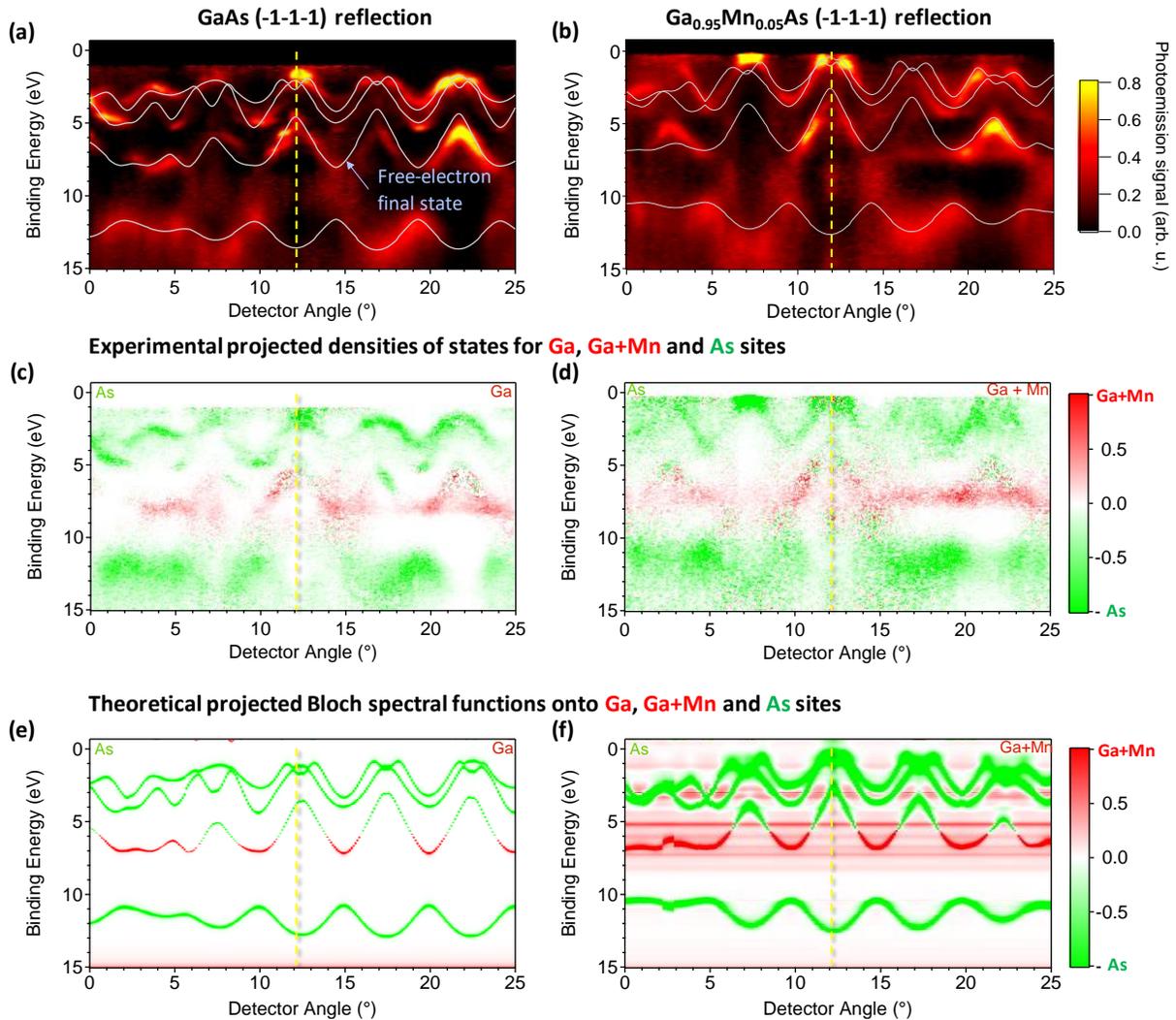

**Figure 5:** Elemental projection of angle-integrated experimental data. Projections represent roughly the matrix-element weighted density of states of element-resolved components. Experimental data for (a) GaAs and (b) (Ga,Mn)As in a (-1-1-1) reflection geometry, with photon energy of ~2.7 keV and integrated over the whole acceptance angle of analyzer are shown as empty circles (red for Ga+Mn) and squares (green for As). The solid curves show element-projected calculated partial densities of states of Ga, As or Ga+Mn, sites, including coherent potential approximation for the doped sample. The blue curve (whose intensity is multiplied by 2) shows the theoretical projection onto Mn only.

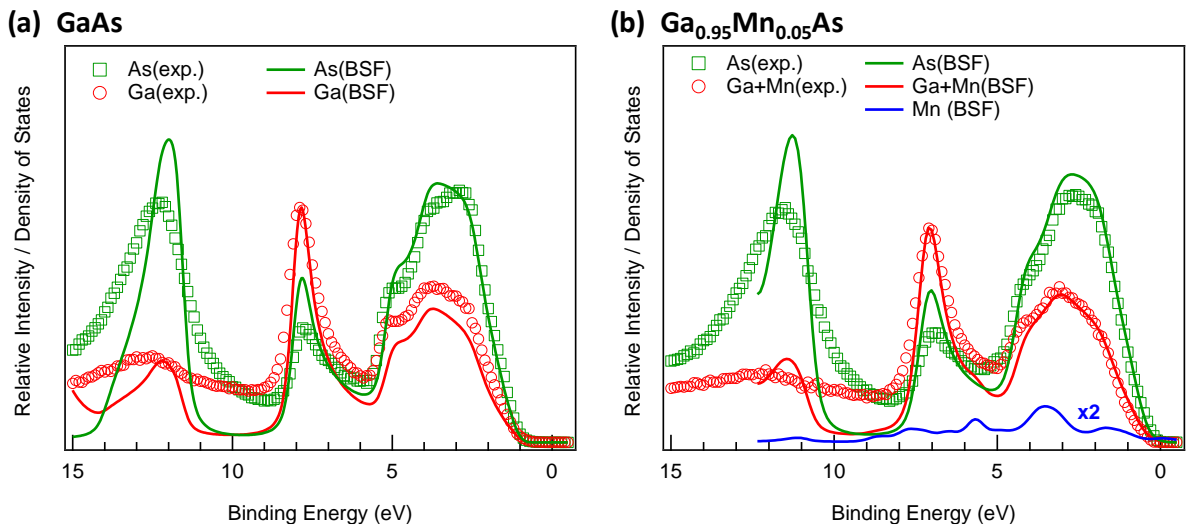